\documentclass[prl,aps,twocolumn,superscriptaddress,showpacs,longbibliography]{revtex4-1}
\usepackage{amsfonts}
\usepackage{amsmath,amssymb}
\usepackage[utf8]{inputenc}
\usepackage{bm}
\usepackage{graphicx}
\usepackage{color}
\usepackage{hyperref}
\hypersetup{colorlinks=true,citecolor={blue},linkcolor={blue},urlcolor={blue}}

\newcommand{\mi}{\mathrm{i}}
\newcommand{\bs}[1]{\mathbb{S}_{#1}}

\begin{document}

\title{Second-order correlations in an exciton-polariton Rabi oscillator}

\author{Yuri G. Rubo}
 \affiliation{Instituto de Energ\'{\i}as Renovables, Universidad Nacional
 Aut\'onoma de M\'exico, Temixco, Morelos, 62580, Mexico}
\author{Alexandra Sheremet}
 \affiliation{Russian Quantum Center, Novaya 100, 143025 Skolkovo, Moscow Region, Russia}
 \affiliation{Department of Theoretical Physics,
 St-Petersburg State Polytechnic University, 195251, St.-Petersburg, Russia}
\author{Alexey Kavokin}
 \affiliation{Russian Quantum Center, Novaya 100, 143025 Skolkovo, Moscow Region, Russia}
 \affiliation{School of Physics and Astronomy, University of Southampton, SO17 1BJ, Southampton,
 United Kingdom}
 \affiliation{Spin Optics Laboratory, St.-Petersburg State University, 198504,
 Peterhof, St.-Petersburg, Russia}

\begin{abstract}
We develop the theoretical formalism to calculate second-order correlations in dissipative
exciton-polariton system and we propose intensity-intensity correlation experiments to reveal the
physics of exciton-light coupling in semiconductor microcavities in the Rabi oscillation regime.
We predict a counter-intuitive behaviour of the correlator between upper and lower polariton
branches: due to the decoherence caused by stochastic exciton-photon conversions this correlator
is expected to decrease below 1, while the individual second-order coherence of upper and lower
polaritons exhibits non-monotonous bunching.
\end{abstract}

\date{\today}

\pacs{71.36.+c, 42.50.Ar, 42.50.Md, 78.47.J-}

\maketitle

\begin{figure}[t]
\includegraphics[width=0.99\columnwidth]{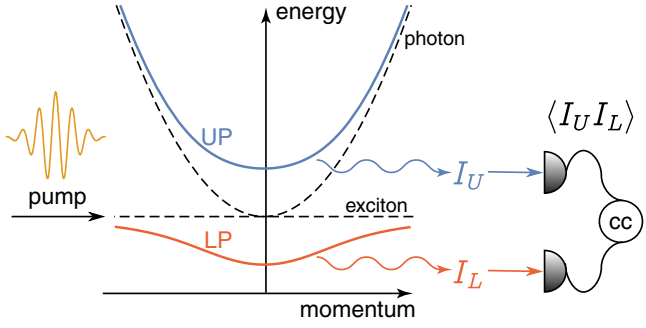}
\caption{(Color online) The scheme of the two-color intensity-intensity correlation measurement
 between upper and lower polariton branches. A short pulse of light excites the system and induces
 polariton Rabi oscillations. Correlations of the intensities of secondary emission harmonics
 corresponding to upper and lower polariton branches are studied.}
\label{fig-setup}
\end{figure}

\emph{Introduction.---}Recent decades are manifested by a tremendous progress in the physics of
strongly-coupled light-matter systems. In particular, semiconductor microcavity \cite{kavokin07}
structures have demonstrated a number of fascinating coherent many-body effects, e.g., polariton
lasing \cite{christopoulos07} and formation of Bose-Einstein condensates of exciton-polaritons
\cite{kasprzak06}. Exciton-polaritons may be viewed as quantum superposition states of light and
matter \cite{carusotto13}. Their dual nature is visualised in polariton Rabi oscillations: beats
between exciton and photon states in semiconductor microcavities in the strong coupling regime
\cite{norris94}. Usually, the Rabi oscillations are excited by a short laser pulse which
simultaneously and equally populates the upper polariton (UP) and the lower polariton (LP)
branches. Initially, the system is in a purely photonic state. Then, due to the splitting between
UP and LP branches, it starts developing the excitonic component, becomes purely excitonic after
several fractions of a picosecond, then returns to the photonic state, and so on. Polariton
Rabi-oscillations have been experimentally observed by many groups
\cite{brunetti06,degiorgi14,dominici14}. Usually, only several periods of oscillations could be
resolved in these studies. The magnitude of oscillations was found to decrease with time and
eventually vanish due to some decoherence processes. The nature of these processes still needs to
be revealed. One process is the phonon-assisted scattering between UP and LP branches that leads to
depopulation of the upper branch and breaks coherence between two branches \cite{martin07}. In the
recent work \cite{kavokin15} we have addressed a process of stochastic exciton-photon conversion,
which is crucial in the weak exciton-photon coupling regime and possibly plays an important role in
the strong coupling regime too. The question of existence of stochastic processes even in the
strong coupling regime is important for understanding the quantum properties of exciton-polaritons.
In which extent polaritons can be considered as coherent superpositions of photons and excitons?
How accurate would be the description of a polariton gas in terms of an exciton-photon mixture? The
dynamics of exciton-photon correlators in the polariton lasing regime calculated in our previous
paper \cite{kavokin15} would help answering these questions. Unfortunately, direct experimental
measurements of exciton-photon correlations seem quite tricky. Here we show that instead of
counting individual excitons and photons one can access the stochastic exciton-photon
transformation kinetics by doing a purely optical and much simpler experiment. Namely, one can
study two-color intensity-intensity correlations of light emitted from UP and LP branches in the
strong coupling regime. The scheme of proposed optical experiment is shown in Fig.\
\ref{fig-setup}.

The finite lifetime of excitons and photons does not destroy the coherence of Rabi oscillations,
only the amplitude of the signal decays in time after pulsed excitation of the system. The
stochastic conversion of particles, however, has a profound destructing effect on the Rabi process.
The result is qualitatively similar to the suppression of the tunneling for a particle in a
double-well potential in the presence of dissipation \cite{caldeira83}. In this Letter, we present
the exact theory of second-order correlations on exciton-polariton system in the presence of
dissipation due to exciton-photon conversion. We predict a counterintuitive variation of the
intensity-intensity correlator between UP and LP branches (two-color correlation experiment) and
show that this variation may be considered as a signature of stochastic conversions in the strong
coupling regime.

From the point of view of quantum statistics, the loss of coherence in a light source is manifested
by the variation of the intensity-intensity correlator $g^{(2)}$ that ranges between 1 (purely
coherent light) and 2 (thermal distribution of photons) for classical sources \cite{mandel95}. In
the regime of polariton Rabi oscillations, the quantum coherence between two branches can be
characterised by a similar quantity $g_{ul}$ defining the correlator of intensities of light
emitted from the UP and LP branches. A coherent optical excitation sets $g_{ul}=1$ initially and
simple depopulation of the branches does not affect this value. However, as we show below, the
stochastic exciton-photon conversion processes manifest themselves in deviation of the UP-LP
intensity correlator from unity, so that it becomes lower than 1 at large times.

\emph{Theoretical model.---}The dissipative exciton-polariton system can be described by the
Liouville equation for the full density matrix \cite{kavokin15}
\begin{equation}\label{DensMat}
 \frac{d\hat{\rho}(t)}{dt}=\frac{\mi}{\hbar}[\hat{\rho},\hat{H}_{0}]-\sum_{j}\frac{g_j}{2}
 \left( [\hat{A}_{j}^{\dagger},\hat{A}_{j}\hat{\rho}]
 + [\hat{\rho}\hat{A}_{j}^{\dagger},\hat{A}_{j}] \right).
\end{equation}
Here the Hamiltonian is
\begin{equation}\label{H0}
 \hat{H}_0=\frac{\hbar}{2}\left\{
    \Delta(\hat{b}^\dagger\hat{b}-\hat{a}^\dagger\hat{a})
   +\omega_R(\hat{a}^\dagger\hat{b}+\hat{b}^\dagger\hat{a})
 \right\},
\end{equation}
where $\hat{a}$ and $\hat{b}$ are the exciton and photon annihilation operators, respectively,
$\Delta$ is the exciton-photon detuning, and $\omega_{R}$ is the exciton-photon coupling frequency.
In Eq.\ \eqref{DensMat} we introduced three ($j=x,c,u$) single-particle Lindblad terms with
$\hat{A}_{x}=\hat{a}$, $\hat{A}_{c}=\hat{b}$, and $\hat{A}_{u}=(\hat{a}+\hat{b})$. The first two
terms describe the direct depopulation of the exciton and photon states. The third term with
coefficient $g_{u}\equiv\gamma^{\prime}$ simply renormalizes the exciton and the cavity photon
lifetimes, $\tau_{x}=(g_{x}+\gamma^{\prime})^{-1}$ and $\tau_{c}=(g_{c}+\gamma^{\prime})^{-1}$, on
the one hand. On the other hand, it models the cross-relaxation and allows phenomenologically for
an additional decay from the upper polariton state that is crucial for description of realistic
microcavities. Finally, there are two ($j=xc,cx$) terms describing the conversion processes, with
$\tau_{xc}=\tau_{cx}=g_{xc}^{-1}$ being the exciton-photon conversion time and operators
$\hat{A}_{xc}=\hat{b}^{\dagger }\hat{a}$ and $\hat{A}_{cx}=\hat{a}^{\dagger }\hat{b}$.

In what follows we shall consider different first- and second-order correlations. For the density
matrix satisfying \eqref{DensMat} it is possible to find the closed systems of equations for the
correlators of any order. To proceed, it is convenient to introduce the spin operators
$\hat{s}_{\mu}$ with $\mu=0,1,2,3$:
\begin{subequations}
 \label{opers}
\begin{eqnarray}
  && \hat{s}_0 
               =\frac{1}{2}\big( \hat{a}^\dagger\hat{a} + \hat{b}^\dagger\hat{b} \big), \\
  && \hat{s}_1 
               =\frac{1}{2}\big( \hat{a}^\dagger\hat{a}-\hat{b}^\dagger\hat{b} \big), \\
  && \hat{s}_2 
               =\frac{\mi}{2}\big( \hat{a}^\dagger\hat{b}-\hat{b}^\dagger\hat{a} \big),  \\
  && \hat{s}_3 
               =\frac{1}{2}\big( \hat{a}^\dagger\hat{b}+\hat{b}^\dagger\hat{a} \big).
\end{eqnarray}
\end{subequations}
The intensities of light emitted from UP and LP branches can be found from averages
$S_{\mu}(t)=\langle \hat{s}_{\mu}\rangle$, which satisfy
\begin{subequations}
 \label{singleS}
\begin{eqnarray}
  && \dot{S}_0 = -\Gamma S_0 + \gamma S_1 - \gamma^\prime S_3, \\
  && \dot{S}_1 = -(\Gamma+2\tau_{xc}^{-1})S_1 + \gamma S_0 - \omega_R S_2, \\
  && \dot{S}_2 = -(\Gamma+\tau_{xc}^{-1})S_2 + \Delta S_3 + \omega_R S_1, \\
  && \dot{S}_3 = -(\Gamma+\tau_{xc}^{-1})S_3 - \gamma^\prime S_0 - \Delta S_2.
\end{eqnarray}
\end{subequations}
Here $\Gamma =(g_{x}+g_{c}+2\gamma ^{\prime })/2$ and $\gamma =(g_{c}-g_{x})/2$.
%

To compute the second-order coherence it is convenient to operate with the averages of the normal
ordered products $\bs{\mu\nu}(t)=\langle :\!\!\hat{s}_{\mu}\hat{s}_{\nu}\!\!:\rangle$ with
$\mu,\nu=0,1,2,3$. The components of the symmetric tensor $\bs{\mu\nu}(t)$ obey the equations
\begin{widetext}
\begin{subequations}
 \label{doubleS}
\begin{eqnarray}
  && \dot{\mathbb{S}}_{11} =
     -2\Gamma\bs{11}-2\tau_{xc}^{-1}(2\bs{11}-\bs{22}-\bs{33})+2\gamma\bs{01}-2\omega_R\bs{12}, \\
  && \dot{\mathbb{S}}_{22} =
     -2\Gamma\bs{22}-2\tau_{xc}^{-1}(\bs{22}-\bs{11})+2\Delta\bs{23}+2\omega_R\bs{12}, \\
  && \dot{\mathbb{S}}_{33} =
     -2\Gamma\bs{33}-2\tau_{xc}^{-1}(\bs{33}-\bs{11})-2\gamma^\prime\bs{03}-2\Delta\bs{23}, \\
  && \dot{\mathbb{S}}_{01} =
     -2\Gamma\bs{01}-2\tau_{xc}^{-1}\bs{01}+\gamma(\bs{00}+\bs{11})-\gamma^\prime\bs{13}-\omega_R\bs{02}, \\
  && \dot{\mathbb{S}}_{02} =
     -2\Gamma\bs{02}-\tau_{xc}^{-1}\bs{02}+\gamma\bs{12}-\gamma^\prime\bs{23}+\Delta\bs{03}+\omega_R\bs{01}, \\
  && \dot{\mathbb{S}}_{03} =
     -2\Gamma\bs{03}-\tau_{xc}^{-1}\bs{03}+\gamma\bs{13}-\gamma^\prime(\bs{00}+\bs{33})-\Delta\bs{02}, \\
  && \dot{\mathbb{S}}_{12} =
     -2\Gamma\bs{12}-5\tau_{xc}^{-1}\bs{12}+\gamma\bs{02}+\Delta\bs{13}+\omega_R(\bs{11}-\bs{22}), \\
  && \dot{\mathbb{S}}_{13} =
     -2\Gamma\bs{13}-5\tau_{xc}^{-1}\bs{13}+\gamma\bs{03}-\gamma^\prime\bs{01}-\Delta\bs{12}-\omega_R\bs{23}, \\
  && \dot{\mathbb{S}}_{23} =
     -2\Gamma\bs{23}-2\tau_{xc}^{-1}\bs{23}-\gamma^\prime\bs{02}+\Delta(\bs{33}-\bs{22})+\omega_R\bs{13}.
\end{eqnarray}
\end{subequations}
\end{widetext}
Here we omitted the equation for $\bs{00}$, which follows from the identity
$\bs{00}=\bs{11}+\bs{22}+\bs{33}$, namely,
$\dot{\mathbb{S}}_{00}=-2\Gamma\bs{00}+2\gamma\bs{01}-2\gamma^{\prime}\bs{03}$. It should be noted
that the Eqs.\ (\ref{singleS}a-d) and (\ref{doubleS}a-i) are exact consequences of the quantum
Liouville equation for the density matrix (\ref{DensMat}). No quasiclassical assumptions have been
made, and one can use these equations for the analysis of evolution of the quantum states of the
Rabi oscillator.

In what follows, we consider the pulsed excitation of a photonic state assuming that an ultrashort
pulse of light arrives at $t=0$. For the case of initial coherent photonic state with $N$ photons
in average, the initial condition to Eqs.\ (\ref{singleS}a-d) are $S_0(0)=-S_1(0)=N/2$  and
$S_2(0)=S_3(0)=0$. For Eqs.\ (\ref{doubleS}a-i) we have $\bs{00}=\bs{11}=-\bs{01}=N^{2}/4$, and the
other components of the tensor are zero. In the exotic reference case of an initial Fock state with
$N$ photons, the latter should be changed to $\bs{00}=\bs{11}=-\bs{01}=N(N-1)/4$.

\begin{figure}[t]
\includegraphics[width=0.99\columnwidth]{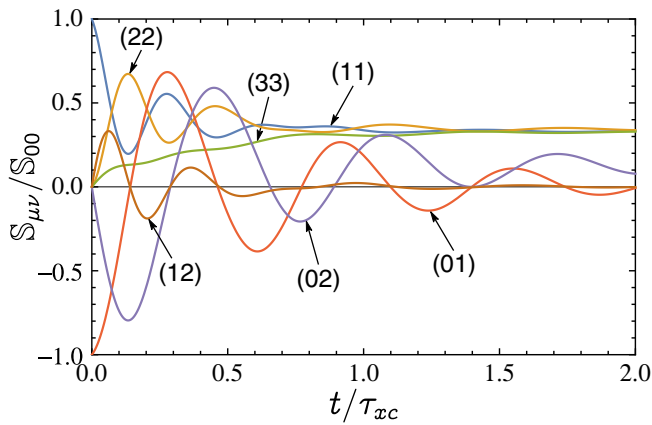}
\caption{(Color online) Showing the time dependencies of normalized non-zero
 components of the tensor $\mathbb{S}_{\mu\nu}$ after excitation of a photon state at $t=0$.
 The curves are labeled by indices $(\mu\nu)$. The parameters are $\tau_c/\tau_{xc}=0.4$,
 $\tau_x/\tau_{xc}=1.6$, $\omega_R\tau_{xc}=10$, and $\Delta=\gamma^\prime=0$.}
\label{fig-tensor}
\end{figure}

In the most experimentally relevant case of photonic excitation the components $S_{0,1,2}(t)$
exhibit decaying Rabi oscillations, and in the absence of stochastic processes ($\tau_{xc}^{-1}=0$)
one has $S_0^{2}=S_{1}^{2}+S_{2}^{2}+S_{3}^{2}$. This equality is violated in the case of a finite
stochastic conversion time $\tau_{xc}$, since the exciton-photon conversion results in decoherence.
The evolution of tensor $\bs{\mu\nu}$ is more complex and it is characterized by appearance of
finite off-diagonal correlations at long times. Even in the simplest case of $\gamma^{\prime}=0$
and zero detuning, which is shown in Fig.\ \ref{fig-tensor}, there are six non-zero components. If
exciton and photon decay rates are slow compared to the decoherence rate
($\tau_{x},\tau_{c}\gg\tau_{xc}$), the diagonal $\bs{11}$, $\bs{22}$, and $\bs{33}$ components tend
to $\bs{00}/3$ at long times, indicating the equidistribution of populations of excitons and
photons. This tendency is only approximately observed if decay and decoherence rates are of the
same order. For the parameters in Fig.\ \ref{fig-tensor}, we have at long times
$\bs{11}\rightarrow0.331\bs{00}$, $\bs{22}\rightarrow0.342\bs{00}$, and $\bs{33}\rightarrow
0.327\bs{00}$. Also, at long times, there appear finite off-diagonal correlations,
$\bs{01}\rightarrow0.0124\bs{00}$, $\bs{02}\rightarrow0.122\bs{00}$, and
$\bs{12}\rightarrow0.00147\bs{00}$.

\begin{figure}[tb]
\includegraphics[width=0.99\columnwidth]{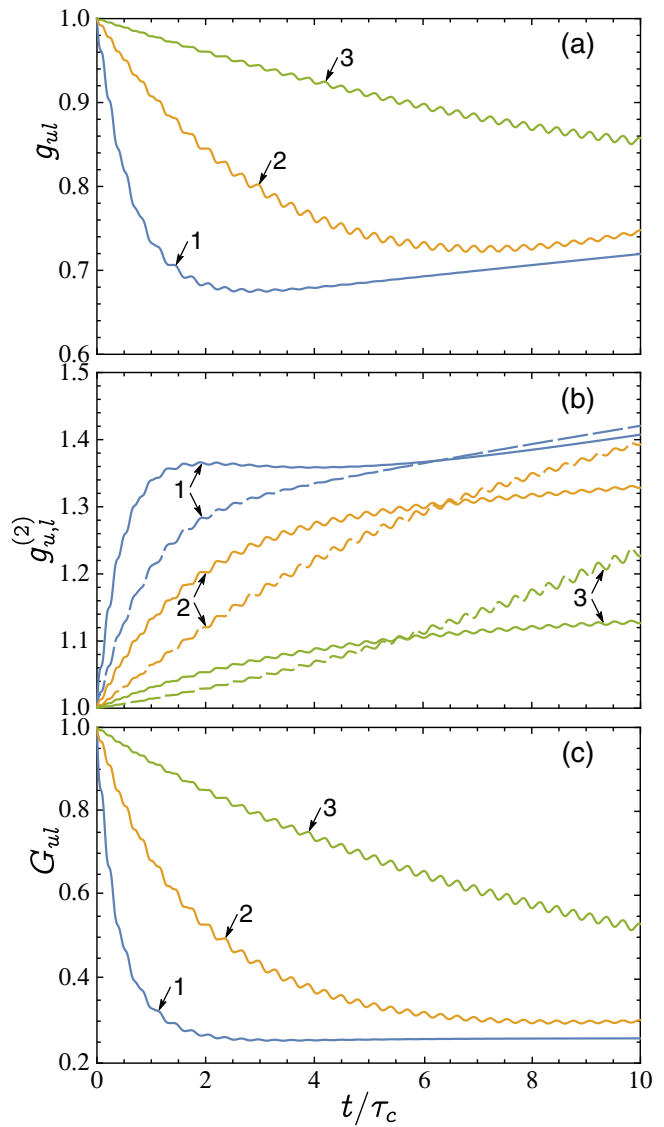}
\caption{(Color online) The time evolution of the correlators
 (a) $g_{ul}$, (b) $g_{u}^{(2)}$ (solid lines) and $g_{l}^{(2)}$ (dashed lines), (c) $G_{ul}$
 for different exciton-photon conversion times,
 (1) $\tau_c/\tau_{xc} = 0.5$, (2) $\tau_c/\tau_{xc} = 0.1$, and (3) $\tau_c/\tau_{xc} = 0.02$.
 The other parameters are $\tau_x = 4\tau_c$, $\omega_{R}\tau_c = 20$, $\Delta\tau_c = -5$,
 and $\gamma^\prime=0$. }
\label{fig-corrs}
\end{figure}

\emph{Intensity-intensity correlations.---}The Hamiltonian \eqref{H0} is diagonalized in the basis
of LP and UP states with the annihilation operators $\hat{c}_{l}=\hat{a}\cos\phi-\hat{b}\sin\phi$
and $\hat{c}_{u}=\hat{a}\sin\phi +\hat{b}\cos\phi$, where the auxiliary angle $\phi$ is defined by
$\tan(2\phi )=\omega_{R}/\Delta$. The polariton occupation numbers are
\begin{equation}
 n_{i}=\langle \hat{c}_{i}^{\dagger }\hat{c}_{i}\rangle
 = S_{0} \mp S_{1}\cos(2\phi) \pm S_{3}\sin(2\phi),
 \label{npol}
\end{equation}
where the upper and the lower signs correspond to the UP ($i=u$) and the LP ($i=l$) branches,
respectively.

There are several second-order correlators that can be experimentally accessed for an
exciton-polariton Rabi oscillator. Those of major interest are the UP-LP intensities correlator
$g_{ul}(t)$, which can be determined as
\begin{multline}
 g_{ul}=\frac{\langle \hat{c}_{u}^{\dagger}\hat{c}_{l}^{\dagger}\hat{c}_{u}\hat{c}_{l} \rangle}%
 {\langle\hat{c}_{u}^{\dagger}\hat{c}_{u}\rangle\langle\hat{c}_{l}^{\dagger}\hat{c}_{l}\rangle} \\
 =\frac{\bs{11}\sin ^{2}(2\phi)+\bs{22}+\bs{33}\cos^{2}(2\phi)+\bs{13}\sin(4\phi )}%
 {S_{0}^{2}-[S_{1}\cos (2\phi)-S_{3}\sin (2\phi )]^{2}},
 \label{gult}
\end{multline}
the second-order coherence for UP and LP,
\begin{multline}
 g_{i}^{(2)}=n_{i}^{-2}\langle\hat{c}_{i}^{\dagger}\hat{c}_{i}^{\dagger}\hat{c}_{i}\hat{c}_{i}\rangle \\
 = n_{i}^{-2}\left[ \bs{00}+\bs{11}\cos ^{2}(2\phi )+\bs{33}\sin ^{2}(2\phi )\right.  \\
             \left. \mp 2\bs{01}\cos (2\phi )\pm 2\bs{03}\sin (2\phi )-\bs{13}\sin (4\phi)\right],
 \label{gsame2}
\end{multline}
with the same convention about the signs as in Eq.\ \eqref{npol}, and the generalized UP-LP
correlator expressed trough the previous three correlators,
\begin{equation}
 G_{ul}=\frac{g_{ul}^{2}}{g_{u}^{(2)}g_{l}^{(2)}}
 =\frac{\langle \hat{c}_{u}^{\dagger}\hat{c}_{l}^{\dagger}\hat{c}_{u}\hat{c}_{l}\rangle^{2}}%
 { \langle \hat{c}_{u}^{\dagger}\hat{c}_{u}^{\dagger}\hat{c}_{u}\hat{c}_{u}\rangle
   \langle \hat{c}_{l}^{\dagger}\hat{c}_{l}^{\dagger}\hat{c}_{l}\hat{c}_{l}\rangle }.
 \label{Bigul}
\end{equation}

The above correlators equate to unity for coherent Rabi oscillations, and their deviation from 1 is
a smoking gun of the stochastic exciton-photon conversion processes. Moreover, the effect of
decoherence is well pronounced even for rather long conversion times $\tau_{xc}$, as it is seen in
Fig.\ \ref{fig-corrs}(a-c). The UP-LP correlator \eqref{gult} shown in Fig.\ \ref{fig-corrs}(a)
decreases below 1, and for short enough $\tau_{xc}$ can become close to 2/3. This happens because
the exciton-photon conversion process tend to form the equidistribution of particles, and the
perfect equidistribution gives $g_{ul}=2/3$ \cite{kavokin15}. However, the equidistribution does
not hold with time, because the excitons and photons are removed from the cavity with different
rates $\tau_x^{-1}\ne\tau_c^{-1}$. As a result, the time dependence of $g_{ul}$ is nonmonotonous
and there is a slow growth of this correlator at long times. The evolution of the polariton
second-order coherence \eqref{gsame2} is also in general nonmonotonous and exhibits bunching,
$g_{u,l}^{(2)}>1$, see Fig.\ \ref{fig-corrs}(b). Since both $g_{ul}$ and $g_{u,l}^{(2)}$ increase
at long times, it is convenient to consider the combined correlator $G_{ul}$ \eqref{Bigul} which
tends to a constant value at $t\rightarrow\infty$. This saturation value is 1/4 for short
$\tau_{xc}$ \cite{kavokin15}. When $\tau_{xc}$ is comparable or larger than $\tau_c$ the saturation
value is bigger than 1/4 and it depends on $\Delta$ and $\gamma^\prime$. All correlators shown in
Fig.\ \ref{fig-corrs}(a-c) reflect the presence of Rabi oscillations in the system.

\emph{In conclusion,} two-color intensity-intensity correlation measurements between UP and LP
branches in the regime of Rabi-oscillations are expected to shed light onto decoherence caused by
stochastic exciton-photon conversion processes in the system. The two-color correlator is predicted
to go below 1 if the stochastic processes are important. This prediction allows for a relatively
simple experimental test of the decoherence of polariton Rabi oscillator and verification of the
quantum superposition nature of an exciton-polariton state in the strong coupling regime. Finally,
we note that a straightforward extension of our formalism permits evaluation of the noise spectra
of UP and LP emission intensities. These spectra are also expected to be sensitive to the
stochastic exciton-photon conversion.

\bibliography{ulbib}

\begin{thebibliography}{12}%
\makeatletter
\providecommand \@ifxundefined [1]{%
 \@ifx{#1\undefined}
}%
\providecommand \@ifnum [1]{%
 \ifnum #1\expandafter \@firstoftwo
 \else \expandafter \@secondoftwo
 \fi
}%
\providecommand \@ifx [1]{%
 \ifx #1\expandafter \@firstoftwo
 \else \expandafter \@secondoftwo
 \fi
}%
\providecommand \natexlab [1]{#1}%
\providecommand \enquote  [1]{``#1''}%
\providecommand \bibnamefont  [1]{#1}%
\providecommand \bibfnamefont [1]{#1}%
\providecommand \citenamefont [1]{#1}%
\providecommand \href@noop [0]{\@secondoftwo}%
\providecommand \href [0]{\begingroup \@sanitize@url \@href}%
\providecommand \@href[1]{\@@startlink{#1}\@@href}%
\providecommand \@@href[1]{\endgroup#1\@@endlink}%
\providecommand \@sanitize@url [0]{\catcode `\\12\catcode `\$12\catcode
  `\&12\catcode `\#12\catcode `\^12\catcode `\_12\catcode `\%12\relax}%
\providecommand \@@startlink[1]{}%
\providecommand \@@endlink[0]{}%
\providecommand \url  [0]{\begingroup\@sanitize@url \@url }%
\providecommand \@url [1]{\endgroup\@href {#1}{\urlprefix }}%
\providecommand \urlprefix  [0]{URL }%
\providecommand \Eprint [0]{\href }%
\providecommand \doibase [0]{http://dx.doi.org/}%
\providecommand \selectlanguage [0]{\@gobble}%
\providecommand \bibinfo  [0]{\@secondoftwo}%
\providecommand \bibfield  [0]{\@secondoftwo}%
\providecommand \translation [1]{[#1]}%
\providecommand \BibitemOpen [0]{}%
\providecommand \bibitemStop [0]{}%
\providecommand \bibitemNoStop [0]{.\EOS\space}%
\providecommand \EOS [0]{\spacefactor3000\relax}%
\providecommand \BibitemShut  [1]{\csname bibitem#1\endcsname}%
\let\auto@bib@innerbib\@empty
\bibitem [{\citenamefont {Kavokin}\ \emph {et~al.}(2007)\citenamefont
  {Kavokin}, \citenamefont {Baumberg}, \citenamefont {Malpuech},\ and\
  \citenamefont {Laussy}}]{kavokin07}%
  \BibitemOpen
  \bibfield  {author} {\bibinfo {author} {\bibfnamefont {Alexey~V.}\
  \bibnamefont {Kavokin}}, \bibinfo {author} {\bibfnamefont {Jeremy~J.}\
  \bibnamefont {Baumberg}}, \bibinfo {author} {\bibfnamefont {Guillaume}\
  \bibnamefont {Malpuech}}, \ and\ \bibinfo {author} {\bibfnamefont
  {Fabrice~P.}\ \bibnamefont {Laussy}},\ }\href@noop {} {\emph {\bibinfo
  {title} {Microcavities}}}\ (\bibinfo  {publisher} {Oxford University Press},\
  \bibinfo {address} {Oxford},\ \bibinfo {year} {2007})\BibitemShut {NoStop}%
\bibitem [{\citenamefont {Christopoulos}\ \emph {et~al.}(2007)\citenamefont
  {Christopoulos}, \citenamefont {Höger~von Högersthal}, \citenamefont
  {Grundy}, \citenamefont {Lagoudakis}, \citenamefont {Kavokin}, \citenamefont
  {Baumberg}, \citenamefont {Christmann}, \citenamefont {Butté}, \citenamefont
  {Feltin}, \citenamefont {Carlin},\ and\ \citenamefont
  {Grandjean}}]{christopoulos07}%
  \BibitemOpen
  \bibfield  {author} {\bibinfo {author} {\bibfnamefont {S.}~\bibnamefont
  {Christopoulos}}, \bibinfo {author} {\bibfnamefont {G.~Baldassarri}\
  \bibnamefont {Höger~von Högersthal}}, \bibinfo {author} {\bibfnamefont
  {A.~J.~D.}\ \bibnamefont {Grundy}}, \bibinfo {author} {\bibfnamefont {P.~G.}\
  \bibnamefont {Lagoudakis}}, \bibinfo {author} {\bibfnamefont {A.~V.}\
  \bibnamefont {Kavokin}}, \bibinfo {author} {\bibfnamefont {J.~J.}\
  \bibnamefont {Baumberg}}, \bibinfo {author} {\bibfnamefont {G.}~\bibnamefont
  {Christmann}}, \bibinfo {author} {\bibfnamefont {R.}~\bibnamefont {Butté}},
  \bibinfo {author} {\bibfnamefont {E.}~\bibnamefont {Feltin}}, \bibinfo
  {author} {\bibfnamefont {J.-F.}\ \bibnamefont {Carlin}}, \ and\ \bibinfo
  {author} {\bibfnamefont {N.}~\bibnamefont {Grandjean}},\ }\bibfield  {title}
  {\enquote {\bibinfo {title} {Room-temperature polariton lasing in
  semiconductor microcavities},}\ }\href {\doibase 10.1103/PhysRevLett.98.126405} 
  {\bibfield  {journal} {\bibinfo  {journal}
  {Phys. Rev. Lett.}\ }\textbf {\bibinfo {volume} {98}},\ \bibinfo {pages}
  {126405} (\bibinfo {year} {2007})}\BibitemShut {NoStop}%
\bibitem [{\citenamefont {Kasprzak}\ \emph {et~al.}(2006)\citenamefont
  {Kasprzak}, \citenamefont {Richard}, \citenamefont {Kundermann},
  \citenamefont {Baas}, \citenamefont {Jeambrun}, \citenamefont {Keeling},
  \citenamefont {Marchetti}, \citenamefont {Szymańska}, \citenamefont
  {André}, \citenamefont {Staehli}, \citenamefont {Savona}, \citenamefont
  {Littlewood}, \citenamefont {Deveaud},\ and\ \citenamefont
  {Dang}}]{kasprzak06}%
  \BibitemOpen
  \bibfield  {author} {\bibinfo {author} {\bibfnamefont {J.}~\bibnamefont
  {Kasprzak}}, \bibinfo {author} {\bibfnamefont {M.}~\bibnamefont {Richard}},
  \bibinfo {author} {\bibfnamefont {S.}~\bibnamefont {Kundermann}}, \bibinfo
  {author} {\bibfnamefont {A.}~\bibnamefont {Baas}}, \bibinfo {author}
  {\bibfnamefont {P.}~\bibnamefont {Jeambrun}}, \bibinfo {author}
  {\bibfnamefont {J.~M.~J.}\ \bibnamefont {Keeling}}, \bibinfo {author}
  {\bibfnamefont {F.~M.}\ \bibnamefont {Marchetti}}, \bibinfo {author}
  {\bibfnamefont {M.~H.}\ \bibnamefont {Szymańska}}, \bibinfo {author}
  {\bibfnamefont {R.}~\bibnamefont {André}}, \bibinfo {author} {\bibfnamefont
  {J.~L.}\ \bibnamefont {Staehli}}, \bibinfo {author} {\bibfnamefont
  {V.}~\bibnamefont {Savona}}, \bibinfo {author} {\bibfnamefont {P.~B.}\
  \bibnamefont {Littlewood}}, \bibinfo {author} {\bibfnamefont
  {B.}~\bibnamefont {Deveaud}}, \ and\ \bibinfo {author} {\bibfnamefont
  {Le~Si}\ \bibnamefont {Dang}},\ }\bibfield  {title} {\enquote {\bibinfo
  {title} {{B}ose-{E}instein condensation of exciton polaritons},}\ }\href
  {\doibase 10.1038/nature05131} {\bibfield  {journal} {\bibinfo  {journal}
  {Nature (London)}\ }\textbf {\bibinfo {volume} {443}},\ \bibinfo {pages}
  {409--414} (\bibinfo {year} {2006})}\BibitemShut {NoStop}%
\bibitem [{\citenamefont {Carusotto}\ and\ \citenamefont
  {Ciuti}(2013)}]{carusotto13}%
  \BibitemOpen
  \bibfield  {author} {\bibinfo {author} {\bibfnamefont {Iacopo}\ \bibnamefont
  {Carusotto}}\ and\ \bibinfo {author} {\bibfnamefont {Cristiano}\ \bibnamefont
  {Ciuti}},\ }\bibfield  {title} {\enquote {\bibinfo {title} {Quantum fluids of
  light},}\ }\href {\doibase 10.1103/RevModPhys.85.299} {\bibfield  {journal}
  {\bibinfo  {journal} {Rev. Mod. Phys.}\ }\textbf {\bibinfo {volume} {85}},\
  \bibinfo {pages} {299--366} (\bibinfo {year} {2013})}\BibitemShut {NoStop}%
\bibitem [{\citenamefont {Norris}\ \emph {et~al.}(1994)\citenamefont {Norris},
  \citenamefont {Rhee}, \citenamefont {Sung}, \citenamefont {Arakawa},
  \citenamefont {Nishioka},\ and\ \citenamefont {Weisbuch}}]{norris94}%
  \BibitemOpen
  \bibfield  {author} {\bibinfo {author} {\bibfnamefont {T.~B.}\ \bibnamefont
  {Norris}}, \bibinfo {author} {\bibfnamefont {J.-K.}\ \bibnamefont {Rhee}},
  \bibinfo {author} {\bibfnamefont {C.-Y.}\ \bibnamefont {Sung}}, \bibinfo
  {author} {\bibfnamefont {Y.}~\bibnamefont {Arakawa}}, \bibinfo {author}
  {\bibfnamefont {M.}~\bibnamefont {Nishioka}}, \ and\ \bibinfo {author}
  {\bibfnamefont {C.}~\bibnamefont {Weisbuch}},\ }\bibfield  {title} {\enquote
  {\bibinfo {title} {Time-resolved vacuum {R}abi oscillations in a
  semiconductor quantum microcavity},}\ }\href {\doibase 10.1103/PhysRevB.50.14663} 
  {\bibfield  {journal} {\bibinfo  {journal} {Phys. Rev. B}\ }\textbf {\bibinfo {volume} {50}},\ \bibinfo {pages} {14663--14666}
  (\bibinfo {year} {1994})}\BibitemShut {NoStop}%
\bibitem [{\citenamefont {Brunetti}\ \emph {et~al.}(2006)\citenamefont
  {Brunetti}, \citenamefont {Vladimirova}, \citenamefont {Scalbert},
  \citenamefont {Nawrocki}, \citenamefont {Kavokin}, \citenamefont {Shelykh},\
  and\ \citenamefont {Bloch}}]{brunetti06}%
  \BibitemOpen
  \bibfield  {author} {\bibinfo {author} {\bibfnamefont {A.}~\bibnamefont
  {Brunetti}}, \bibinfo {author} {\bibfnamefont {M.}~\bibnamefont
  {Vladimirova}}, \bibinfo {author} {\bibfnamefont {D.}~\bibnamefont
  {Scalbert}}, \bibinfo {author} {\bibfnamefont {M.}~\bibnamefont {Nawrocki}},
  \bibinfo {author} {\bibfnamefont {A.~V.}\ \bibnamefont {Kavokin}}, \bibinfo
  {author} {\bibfnamefont {I.~A.}\ \bibnamefont {Shelykh}}, \ and\ \bibinfo
  {author} {\bibfnamefont {J.}~\bibnamefont {Bloch}},\ }\bibfield  {title}
  {\enquote {\bibinfo {title} {Observation of spin beats at the {R}abi
  frequency in microcavities},}\ }\href {\doibase 10.1103/PhysRevB.74.241101}
  {\bibfield  {journal} {\bibinfo  {journal} {Phys. Rev. B}\ }\textbf {\bibinfo
  {volume} {74}},\ \bibinfo {pages} {241101} (\bibinfo {year}
  {2006})}\BibitemShut {NoStop}%
\bibitem [{\citenamefont {De~Giorgi}\ \emph {et~al.}(2014)\citenamefont
  {De~Giorgi}, \citenamefont {Ballarini}, \citenamefont {Cazzato},
  \citenamefont {Deligeorgis}, \citenamefont {Tsintzos}, \citenamefont
  {Hatzopoulos}, \citenamefont {Savvidis}, \citenamefont {Gigli}, \citenamefont
  {Laussy},\ and\ \citenamefont {Sanvitto}}]{degiorgi14}%
  \BibitemOpen
  \bibfield  {author} {\bibinfo {author} {\bibfnamefont {Milena}\ \bibnamefont
  {De~Giorgi}}, \bibinfo {author} {\bibfnamefont {Dario}\ \bibnamefont
  {Ballarini}}, \bibinfo {author} {\bibfnamefont {Paolo}\ \bibnamefont
  {Cazzato}}, \bibinfo {author} {\bibfnamefont {George}\ \bibnamefont
  {Deligeorgis}}, \bibinfo {author} {\bibfnamefont {Simos~I.}\ \bibnamefont
  {Tsintzos}}, \bibinfo {author} {\bibfnamefont {Zacharias}\ \bibnamefont
  {Hatzopoulos}}, \bibinfo {author} {\bibfnamefont {Pavlos~G.}\ \bibnamefont
  {Savvidis}}, \bibinfo {author} {\bibfnamefont {Giuseppe}\ \bibnamefont
  {Gigli}}, \bibinfo {author} {\bibfnamefont {Fabrice~P.}\ \bibnamefont
  {Laussy}}, \ and\ \bibinfo {author} {\bibfnamefont {Daniele}\ \bibnamefont
  {Sanvitto}},\ }\bibfield  {title} {\enquote {\bibinfo {title} {Relaxation
  oscillations in the formation of a polariton condensate},}\ }\href {\doibase 10.1103/PhysRevLett.112.113602} 
  {\bibfield  {journal} {\bibinfo  {journal}
  {Phys. Rev. Lett.}\ }\textbf {\bibinfo {volume} {112}},\ \bibinfo {pages}
  {113602} (\bibinfo {year} {2014})}\BibitemShut {NoStop}%
\bibitem [{\citenamefont {Dominici}\ \emph {et~al.}(2014)\citenamefont
  {Dominici}, \citenamefont {Colas}, \citenamefont {Donati}, \citenamefont
  {Restrepo~Cuartas}, \citenamefont {De~Giorgi}, \citenamefont {Ballarini},
  \citenamefont {Guirales}, \citenamefont {López~Carreño}, \citenamefont
  {Bramati}, \citenamefont {Gigli}, \citenamefont {del Valle}, \citenamefont
  {Laussy},\ and\ \citenamefont {Sanvitto}}]{dominici14}%
  \BibitemOpen
  \bibfield  {author} {\bibinfo {author} {\bibfnamefont {L.}~\bibnamefont
  {Dominici}}, \bibinfo {author} {\bibfnamefont {D.}~\bibnamefont {Colas}},
  \bibinfo {author} {\bibfnamefont {S.}~\bibnamefont {Donati}}, \bibinfo
  {author} {\bibfnamefont {J.~P.}\ \bibnamefont {Restrepo~Cuartas}}, \bibinfo
  {author} {\bibfnamefont {M.}~\bibnamefont {De~Giorgi}}, \bibinfo {author}
  {\bibfnamefont {D.}~\bibnamefont {Ballarini}}, \bibinfo {author}
  {\bibfnamefont {G.}~\bibnamefont {Guirales}}, \bibinfo {author}
  {\bibfnamefont {J.~C.}\ \bibnamefont {López~Carreño}}, \bibinfo {author}
  {\bibfnamefont {A.}~\bibnamefont {Bramati}}, \bibinfo {author} {\bibfnamefont
  {G.}~\bibnamefont {Gigli}}, \bibinfo {author} {\bibfnamefont
  {E.}~\bibnamefont {del Valle}}, \bibinfo {author} {\bibfnamefont {F.~P.}\
  \bibnamefont {Laussy}}, \ and\ \bibinfo {author} {\bibfnamefont
  {D.}~\bibnamefont {Sanvitto}},\ }\bibfield  {title} {\enquote {\bibinfo
  {title} {Ultrafast control and {R}abi oscillations of polaritons},}\ }\href
  {\doibase 10.1103/PhysRevLett.113.226401} {\bibfield  {journal} {\bibinfo
  {journal} {Phys. Rev. Lett.}\ }\textbf {\bibinfo {volume} {113}},\ \bibinfo
  {pages} {226401} (\bibinfo {year} {2014})}\BibitemShut {NoStop}%
\bibitem [{\citenamefont {Martín}\ \emph {et~al.}(2007)\citenamefont
  {Martín}, \citenamefont {Ballarini}, \citenamefont {Amo}, \citenamefont
  {Viña},\ and\ \citenamefont {André}}]{martin07}%
  \BibitemOpen
  \bibfield  {author} {\bibinfo {author} {\bibfnamefont {M.~D.}\ \bibnamefont
  {Martín}}, \bibinfo {author} {\bibfnamefont {D.}~\bibnamefont {Ballarini}},
  \bibinfo {author} {\bibfnamefont {A.}~\bibnamefont {Amo}}, \bibinfo {author}
  {\bibfnamefont {L.}~\bibnamefont {Viña}}, \ and\ \bibinfo {author}
  {\bibfnamefont {R.}~\bibnamefont {André}},\ }\bibfield  {title} {\enquote
  {\bibinfo {title} {Dynamics of polaritons resonantly created at the upper
  polariton branch},}\ }\href {\doibase 10.1016/j.spmi.2007.03.029} {\bibfield
  {journal} {\bibinfo  {journal} {Superlattices and Microstructures}\ }\textbf
  {\bibinfo {volume} {41}},\ \bibinfo {pages} {328--332} (\bibinfo {year}
  {2007})}\BibitemShut {NoStop}%
\bibitem [{\citenamefont {Kavokin}\ \emph {et~al.}(2015)\citenamefont
  {Kavokin}, \citenamefont {Sheremet}, \citenamefont {Shelykh}, \citenamefont
  {Lagoudakis},\ and\ \citenamefont {Rubo}}]{kavokin15}%
  \BibitemOpen
  \bibfield  {author} {\bibinfo {author} {\bibfnamefont {Alexey~V.}\
  \bibnamefont {Kavokin}}, \bibinfo {author} {\bibfnamefont {Alexandra~S.}\
  \bibnamefont {Sheremet}}, \bibinfo {author} {\bibfnamefont {Ivan~A.}\
  \bibnamefont {Shelykh}}, \bibinfo {author} {\bibfnamefont {Pavlos~G.}\
  \bibnamefont {Lagoudakis}}, \ and\ \bibinfo {author} {\bibfnamefont
  {Yuri~G.}\ \bibnamefont {Rubo}},\ }\bibfield  {title} {\enquote {\bibinfo
  {title} {Exciton-photon correlations in bosonic condensates of
  exciton-polaritons},}\ }\href {\doibase 10.1038/srep12020} {\bibfield
  {journal} {\bibinfo  {journal} {Sci. Rep.}\ }\textbf {\bibinfo {volume}
  {5}},\ \bibinfo {pages} {12020} (\bibinfo {year} {2015})}\BibitemShut
  {NoStop}%
\bibitem [{\citenamefont {Caldeira}\ and\ \citenamefont
  {Leggett}(1983)}]{caldeira83}%
  \BibitemOpen
  \bibfield  {author} {\bibinfo {author} {\bibfnamefont {A.~O.}\ \bibnamefont
  {Caldeira}}\ and\ \bibinfo {author} {\bibfnamefont {A.~J.}\ \bibnamefont
  {Leggett}},\ }\bibfield  {title} {\enquote {\bibinfo {title} {Quantum
  tunnelling in a dissipative system},}\ }\href {\doibase 10.1016/0003-4916(83)90202-6} 
  {\bibfield  {journal} {\bibinfo  {journal}
  {Annals of Physics}\ }\textbf {\bibinfo {volume} {149}},\ \bibinfo {pages}
  {374--456} (\bibinfo {year} {1983})}\BibitemShut {NoStop}%
\bibitem [{\citenamefont {Mandel}\ and\ \citenamefont {Wolf}(1995)}]{mandel95}%
  \BibitemOpen
  \bibfield  {author} {\bibinfo {author} {\bibfnamefont {L.}~\bibnamefont
  {Mandel}}\ and\ \bibinfo {author} {\bibfnamefont {E.}~\bibnamefont {Wolf}},\
  }\href@noop {} {\emph {\bibinfo {title} {Optical Coherence and Quantum
  Optics}}}\ (\bibinfo  {publisher} {Cambridge University},\ \bibinfo {address}
  {Cambridge},\ \bibinfo {year} {1995})\BibitemShut {NoStop}%
\end{thebibliography}%

\end{document}